\documentclass[journal, a4paper]{IEEEtran}

\usepackage{amsfonts}
\usepackage{amsmath}
\usepackage{algorithm}
\usepackage{algorithmic}
\usepackage{cite}
\usepackage{amsmath}
\usepackage{graphicx}
\usepackage{caption}

\newtheorem{define}{Definition}
\begin{document}

\title{Compressive Massive Access for Internet of Things: Cloud Computing or Fog Computing?}
\author{\IEEEauthorblockN{Malong~Ke\IEEEauthorrefmark{1}\IEEEauthorrefmark{2}, Zhen~Gao\IEEEauthorrefmark{1}\IEEEauthorrefmark{2}, and Yongpeng Wu\IEEEauthorrefmark{3}}\\
\IEEEauthorblockA{\IEEEauthorrefmark{1}School of Information and Electronics, Beijing Institute of Technology, Beijing 100081, China}\\
\IEEEauthorblockA{\IEEEauthorrefmark{2}Advanced Research Institute of Multidisciplinary Science, Beijing Institute of Technology, Beijing 100081, China}\\
\IEEEauthorblockA{\IEEEauthorrefmark{3}Department of Electronic Engineering, Shanghai Jiao Tong University, Shanghai 200240, China}\\
Email: kemalong@bit.edu.cn, gaozhen16@bit.edu.cn, yongpeng.wu@sjtu.edu.cn}

\maketitle

\begin{abstract}
This paper considers the support of grant-free massive access and solves the challenge of active user detection and channel estimation in the case of a massive number of users.
By exploiting the sparsity of user activities, the concerned problems are formulated as a compressive sensing problem, whose solution is acquired by approximate message passing (AMP) algorithm.
Considering the cooperation of multiple access points, for the deployment of AMP algorithm, we compare two processing paradigms, cloud computing and fog computing, in terms of their effectiveness in guaranteeing ultra reliable low-latency access.
For cloud computing, the access points are connected in a cloud radio access network (C-RAN) manner, and the signals received at all access points are concentrated and jointly processed in the cloud baseband unit.
While for fog computing, based on fog radio access network (F-RAN), the estimation of user activity and corresponding channels for the whole network is split, and the related processing tasks are performed at the access points and fog processing units in proximity to users.
Compared to the cloud computing paradigm based on traditional C-RAN, simulation results demonstrate the superiority of the proposed fog computing deployment based on F-RAN.
\end{abstract}


\section{Introduction}

With the advent of the Internet-of-Things (IoT) era, massive machine-type communications (mMTC) has been identified as one of the core services in future wireless networks \cite{Dohler_CM'17}.
In this context, the future base stations (BSs) are required to enable connectivity for billions of user equipments.
However, the support of low-latency massive access for massive connectivity is still challenging in current wireless networks \cite{Bockelmann_Access'18}.
Due to the massive number of users, traditional grant-based random access (RA) protocols would suffer from unacceptably high access latency and require extremely complicated collision resolution schemes \cite{{Hasan_CM'13}, {Laya_CST'14}}.
Hence, the grant-free RA protocol is recently proposed as a promising alternative, where each active user directly transmits its pilots and data to the BS without access scheduling in advance \cite{Zhang_CL'16}.
Furthermore, a key characteristic of mMTC is the sporadic traffic of users, i.e., out of many potential users, only a small number are activated and want to access the network in any given time interval \cite{Liu_SPM'18}.
Thus, in grant-free RA, the BS has to detect the active users and estimate their channels, which are vital for the subsequent data detection \cite{Shao_IoTJ'19}.
However, due to the large number of users and the limited channel coherence time, it is not possible to assign orthogonal pilot sequences to all users and the active user detection (AUD) is emerging as a new challenge \cite{Liu_SPM'18}.

By leveraging the sparsity of users' traffic, for grant-free massive access, several compressive sensing (CS)-based methods have been proposed to simultaneously detect active users and their data \cite{{Shim_CL'12}, {Wang_CL'16}, {Jeong_TVT'18}}, while assuming only single-antenna BS and the availability of perfect channel state information.
To jointly perform AUD and channel estimation (CE), the authors in \cite{Liu_TSP'18} developed an approximate message passing (AMP)-based access scheme, where the BS equipped with a massive number of antennas was employed.
On this basis, the virtual angular domain sparsity of massive multi-input multi-output (MIMO) channels was leveraged for further enhanced performance \cite{Ke_TSP'19}.
Moreover, this work was investigated extensively under the cloud radio access network (C-RAN), where two cloud-centric approaches were proposed to support grant-free massive access \cite{{Xu_ICC'15}, {He_ICC'17}}.
In \cite{He_WiOpt'18}, based on fog radio access network (F-RAN), the authors developed a fog computing deployment for AUD and CE, which splits the corresponding computation into multiple units and solves the problem in a distributed manner.
Besides, the pros and cons of cloud computing and fog computing paradigms are analyzed in \cite{Dustdar_SOSE'19}.


In this paper, we investigate massive access under two promising network architectures, C-RAN and F-RAN, where the practical processing deployment for AUD and CE is considered.
Since the IoT users are usually power limited, the access points should be densely distributed to serve a vast area and to save the transmit power of users.
For cloud computing with C-RAN, quantities of remote radio heads (RRHs) are distributed in the network and are connected to the cloud baseband unit (BBU) through fronthual links, as illustrated in Fig. \ref{Fig:C-RAN}.
To reduce the cost of access point deployment, the RRHs are only designed for receiving and transmitting signals, thus the AUD and CE are performed in cloud BBU.
Specifically, by exploiting the sporadic traffic of users, the AUD and CE are formulated as a multiple measurement vector (MMV) CS problem, and a MMV-AMP algorithm is developed to acquire the solution.
Motivated by alleviating the burden on BBU and reducing the network congestion, the fog computing paradigm is further investigated for faster response time, which brings cloud capabilities closer to the edge of the network, i.e., closer to the users.
In fog computing with F-RAN, the RRHs are replaced with fog access points (F-APs) having computation capabilities, and several neighboring F-APs are connected to a fog processing unit for cooperation, as illustrated in Fig. \ref{Fig:F-RAN}.
On this basis, the estimation of user activity and related channels for the whole network is split, and the corresponding processing tasks are performed at the F-APs and fog processing units.


\textit{Notations}: Throughout this paper, scalar variables are denoted by normal-face letters, while boldface lower and upper-case letters denote column vectors and matrices, respectively.
The transpose and complex conjugate operators are denoted by ${( \cdot )^{\rm{T}}}$ and ${( \cdot )^*}$, respectively.
$\left[ {\bf{X}} \right]_{k,m}$ is the $(k,m)$-th element of matrix ${{\bf{X}}}{ \in \mathbb{C}^{K \times M}}$; $\left[ {\bf{X}} \right]_{k,:}$  and $\left[ {\bf{X}} \right]_{:,m}$ are the $k$-th row vector and the $m$-th column vector of matrix ${{\bf{X}}}{ \in \mathbb{C}^{K \times M}}$, respectively.
$\left| {\cal K} \right|_{c}$ is the number of elements in set ${\cal K}$, $\left[ K \right]$ denotes the set $\left\{ {1, \cdots ,K} \right\}$, and $\rm{supp} \left\{  \cdot  \right\}$ is the support set of a vector or a matrix.
Finally, ${\cal C}{\cal N}\left( {x;\mu,v} \right)$ denotes the complex Gaussian distribution of a random variable $x$ with mean $\mu$ and variance $v$.


\section{System Model}
\label{Sec:II}

\begin{figure}[t]
	\centering
	\includegraphics[width=1\columnwidth,keepaspectratio]
    {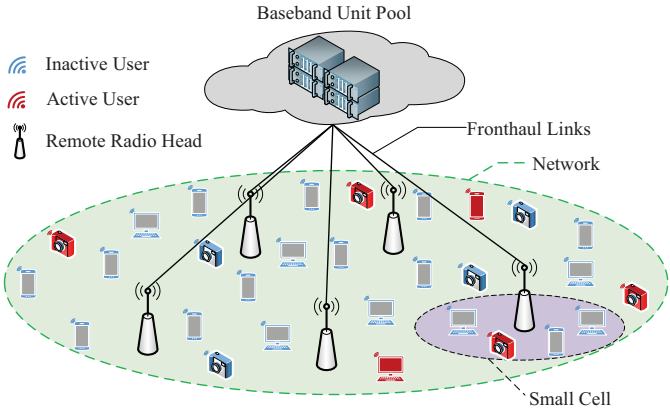}
	\caption{Users exhibit sporadic traffic in massive access, where the C-RAN is employed to cover a vast area and each RRH serves a small cell.}
    \label{Fig:C-RAN}
    \vspace{-5mm}
\end{figure}

\begin{figure}[t]
	\centering
	\includegraphics[width=1\columnwidth,keepaspectratio]
    {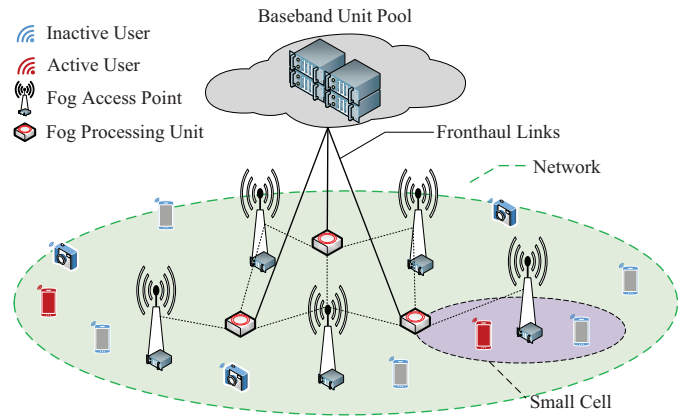}
	\caption{An illustration of F-RAN, with F-APs having the ability to perform processing tasks. Furthermore, every neighboring three F-APs can cooperate in a fog processing unit.}
    \label{Fig:F-RAN}
    \vspace{-3mm}
\end{figure}

Consider a uplink massive access scenario in C-RAN or F-RAN, where the network serves $BK_c$ users distributed in $B$ hexagonal cells, as illustrated in Fig. \ref{Fig:C-RAN} and Fig. \ref{Fig:F-RAN}.
Each cell contains one access point, i.e., RRH or F-AP, located at the center, serving $K_c$ single-antenna users uniformly distributed in its coverage.
Here, the access point is equipped with $M_c$-antenna uniform linear array.
Defining $K = BK_c$, for the $t$-th time slot, the signal ${\bf r}_{b,k}^t \in \mathbb{C}^{M_c \times 1}$ received at the $b$-th access point from the $k$-th user can be expressed as\footnote{Here, $k = (b-1)K_c + c$ with $b \in [B]$ and $c \in [K_c].$}
\begin{equation}\label{Eq:RS_UE_k}
\begin{aligned}
{\bf r}_{b,k}^t = \sqrt{P_k}{\bf h}_{b,k}s_k^t + {\bf n}_b^t,
\end{aligned}
\end{equation}
where $P_k$ denotes the transmit power of the $k$-th user, ${\bf h}_{b,k} \in \mathbb{C}^{M_c \times 1}$ is the channel associated with the $k$-th user and the $b$-th access point, $s_k^t \in {\cal CN}\left(s_k^t; 0, 1\right)$ is the uplink access pilot, and ${\bf n}_b^t$ denotes the additive white Gaussian noise (AWGN).
For typical mMTC, within a given time interval, only a small number of users are activated to access the network.
The user activity indicator is denoted as $\alpha_k$, and is equal to 1 when the $k$-th user is active and 0 otherwise.
Meanwhile, we define the set of active users as ${\cal A} = \left\{k|\alpha_k = 1, 1 \le k \le K\right\}$, and the number of active users is denoted by $K_a = \left|{\cal A}\right|_c$.
Hence, the signal received at the $b$-th access point from all active users is given as follows
\begin{equation}\label{Eq:RS_Slot}
\begin{aligned}
{\bf r}_b^t &= \sum\limits_{k \in {\cal A}} \sqrt{P_k}{\bf h}_{b,k}s_k^t + {\bf n}_b^t\\
            &= \sum\limits_{k=1}^K {\alpha_k}\sqrt{P_k}{\bf h}_{b,k}s_k^t + {\bf n}_b^t.
\end{aligned}
\end{equation}
By considering the large-scale fading and small scale fading, we can model ${\bf h}_{b,k}$ as ${\bf h}_{b,k} = \rho_{b,k}{\widetilde {\bf h}}_{b,k}$, where $\rho_{b,k}$ is the large-scale fading caused by path loss, and ${\widetilde {\bf h}}_{b,k}$ is the small scale fading.
Here, the path loss from the $b$-th access point to the $k$-th user, $\rho_{b,k}$, is given by the standard Log-distance path loss model as $\rho_{b,k} = 128.1 + 37.6{\rm lg}(d_{b,k})$, in which $d_{b,k}$ is the distance measured in km \cite{Chen_TWC'19}.
Furthermore, the small scale fading channel, ${\widetilde {\bf h}}_{b,k}$, is modeled as follows \cite{Ke_TSP'19}
\begin{equation}\label{Eq:Ch_Model}
\begin{aligned}
{\widetilde {\bf h}}_{b,k} = \sum\limits_{l=1}^L\beta_{b,k}^l{\bf a}_{\rm R}\left(\phi_{b,k}^l\right),
\end{aligned}
\end{equation}
where $L$ denotes the number of multi-path components (MPCs), $\beta_{b,k}^l$ is the complex path gain of the $l$-th MPC.
The array response vector ${\bf a}_{\rm R}\left(\phi_{b,k}^l\right)$ is given by ${\bf a}_{\rm R}\left(\phi_{b,k}^l\right) = \left[1, e^{-j2\pi\phi_{b,k}^l}, \cdots, e^{-j2\pi(M-1)\phi_{b,k}^l}\right]$, where $\phi_{b,k}^l = \frac{\widetilde d}\lambda{\rm sin}\left(\varphi _{b,k}^l\right)$.
Here, $\varphi _{b,k}^l$ is the angle of arrival of the $k$-th UE's $l$-th MPC seen from the $b$-th access point, $\lambda$ is the wavelength, and ${\widetilde d} = \lambda/2$ is the antenna spacing.

\section{Active User Detection and Channel Estimation Based on Two Processing Paradigms}

\begin{figure}[t]
	\centering
	\includegraphics[width=1\columnwidth,keepaspectratio]
    {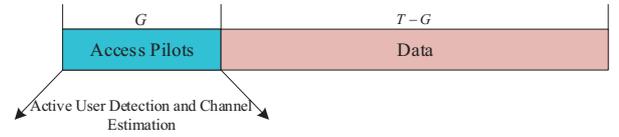}
	\caption{The frame structure of the uplink signals.}
    \label{Fig:Frame_Struc}
    \vspace{-5mm}
\end{figure}

This section details the design of AUD and CE schemes under cloud computing and fog computing paradigms, respectively.
To avoid complicated collision resolutions and the associated latencies, the grant-free random access protocol is adopted, whose frame structure is illustrated in Fig. \ref{Fig:Frame_Struc}.
A frame consists of $T$ time slots, where the first $G$ time slots are used to transmit pilots, and the remaining $(T-G)$ time slots are used for data transmission only.
Here, we assume $T$ is small than the channel coherence time, and the activity of users during the $T$ time slots remains unchanged.
For facilitating the acquisition of user activity and related channel state information, at the access points, the received signals in $G$ successive time slots are collected as
\begin{equation}\label{Eq:RS}
\begin{aligned}
{\bf R}_b = {\bf S}{\bf H}_b + {\bf N}_b, \forall b \in [B],
\end{aligned}
\end{equation}
where ${\bf R}_b = \left[{\bf r}_b^1, {\bf r}_b^2, \cdots,  {\bf r}_b^G\right]^{\rm T} \in \mathbb{C}^{G \times M_c}$, ${\bf S} = \left[{\bf s}^1, {\bf s}^2, \cdots, {\bf s}^G \right]^{\rm T} \in \mathbb{C}^{G \times K}$, ${\bf s}^t = \left[s_1^t, s_2^t, \cdots, s_K^t\right]^{\rm T} \in \mathbb{C}^{K \times 1}$, ${\bf H}_b = \left[{\alpha_1}\sqrt{P_1}{\bf h}_{b,1}, {\alpha_2}\sqrt{P_2}{\bf h}_{b,2}, \cdots, {\alpha_K}\sqrt{P_K}{\bf h}_{b,K}\right]^{\rm T} \in \mathbb{C}^{K \times M_c}$, and ${\bf N}_b = \left[{\bf n}_b^1, {\bf n}_b^2, \cdots, {\bf n}_b^G\right]^{\rm T}$.
In the following, based on received signals ${\bf R}_b$, $\forall b$, and pilot matrix ${\bf S}$, we present the procedure of AUD and CE based on two processing paradigms, respectively.

\subsection{Cloud Computing Based Massive Access}
The procedure of the proposed cloud computing based uplink massive access scheme can be summarized as follows.
\begin{itemize}
\item{\textbf{Step 1:} During the pilot phase of each frame in Fig. \ref{Fig:Frame_Struc}, all active users ${\cal A}$ directly transmit their non-orthogonal access pilot sequences ${\bf s}_k \in \mathbb{C}^{G \times 1}$, $\forall k \in {\cal A}$ to the RRHs.}
\item{\textbf{Step 2:} Each RRH collects the received signals over $G$ successive time slots, and sends the collected signal ${\bf R}_b$ to the cloud via fronthaul links.}
\item{\textbf{Step 3:} Jointly processing the received signals from all RRHs, ${\bf R}_b$, $\forall b$, the BBU pool in the cloud performs AUD and CE for the whole network
                       with known random access pilot matrix  ${\bf S}$.}
\end{itemize}
It is clear that, for cloud computing paradigm, the RRHs close to users are working as relays, and the related processing is performed in the cloud, which is faraway from the users.
At the cloud processing units, the received signals from all RRHs are concentrated as
\begin{equation}\label{Eq:RS_Cloud}
\begin{aligned}
{\bf Y} &= \left[{\bf R}_1, {\bf R}_2, \cdots, {\bf R}_B\right]\\
        &= {\bf S}{\bf X} + {\bf N},
\end{aligned}
\end{equation}
where ${\bf N} = \left[{\bf N}_1, {\bf N}_2, \cdots, {\bf N}_B\right]$ and ${\bf X} \in \mathbb{C}^{K \times M}$ is expressed as
\vspace{-1.5mm}
\begin{equation}\label{Eq:Ch_Mtx_Cloud}
\begin{aligned}
{\bf X} &= \left[{\bf H}_1, {\bf H}_2, \cdots, {\bf H}_B\right]\\
        &= \left[\begin{array}{*{20}{c}} {\alpha_1}{\eta_{1,1}}{{\widetilde {\bf h}_{1,1}}}& \cdots &{\alpha_1}{\eta_{B,1}}{{\widetilde {\bf h}_{B,1}}}\\
                                                           \vdots                & \ddots &                     \vdots \\
                                        {\alpha_K}{\eta_{1,K}}{{\widetilde {\bf h}_{1,K}}}& \cdots &{\alpha_K}{\eta_{B,K}}{{\widetilde {\bf h}_{B,K}}}
           \end{array}\right].
\end{aligned}
\end{equation}
Here, $\eta_{j,k} = \sqrt{P_k}{\rho_{j,k}}$ and $M = BM_c$.

The channel matrix ${\bf X}$ exhibits some individual sparsity properties, which may be useful for achieving ultra reliable low-latency massive access.
Due to the sporadic traffic of users, only a small number of users are active, i.e., $K_a \ll K$ and most of the activity indicators $\alpha_k$ are equal to $0$.
Thus, the channel vector $\left[{\bf X}\right]_{:,m}$ observed at a specific receive antenna of the RRHs, i.e., the $m$-th column of ${\bf X}$, is sparse as
\vspace{-1.5mm}
\begin{equation}\label{Eq:Sparse_UE}
\begin{aligned}
\left|{\rm supp}\left\{ \left[{\bf X}\right]_{:,m} \right\}\right|_c = K_a \ll K.
\end{aligned}
\vspace{-1mm}
\end{equation}
Moreover, given the user activity, i.e., the value of $\alpha_k$, all elements of the $k$-th row of ${\bf X}$ will be zero or non-zero simultaneously, thus all RRH antennas exhibit the same sparsity
\vspace{-1.5mm}
\begin{equation}\label{Eq:Sparse_Str}
\begin{aligned}
{\rm supp}\!\left\{ \left[{\bf X}\right]_{:,1} \right\}\! = {\rm supp}\!\left\{ \left[{\bf X}\right]_{:,2} \right\}\! = \cdots =
{\rm supp}\!\left\{ \left[{\bf X}\right]_{:,M} \right\}.
\end{aligned}
\vspace{-1mm}
\end{equation}

In grant-free based massive access, the active users and the corresponding channels have to be acquired.
These two targets will be simultaneously realized by estimating channel matrix ${\bf X}$ based on the noisy measurements ${\bf Y}$ and pilot matrix ${\bf S}$, where AUD is actually to find the indices of non-zero rows of ${\bf X}$.
Due to the massive numbers of users and limited channel coherence time, it is unlikely to assign orthogonal pilots to each user, thus $G$ is far less than $K$.
This makes estimating ${\bf X}$ based on (\ref{Eq:RS_Cloud}) an under-determined problem, whose solution is hard to be obtained.
By exploiting the sparsity described in (\ref{Eq:Sparse_UE}), the concerned problems can be formulated as a MMV CS problem.
In this paper, a computationally efficient MMV-AMP algorithm is developed for CS recovery, where the sparsity structure in (\ref{Eq:Sparse_Str}) is leveraged to further enhance the recovery performance, as will be detailed in the following.

The MMV-AMP algorithm adopted in this paper belongs to the family of AMP algorithms \cite{Liu_SPM'18}, which is developed under the Bayesian framework.
An intuitive explanation of AMP is that, in the large system limit, i.e., $K \to \infty$, while $\gamma = Ka/K$ and $\kappa = G/K$ are fixed, the matrix estimation problem based on (\ref{Eq:RS_Cloud}) can be approximately decoupled into $KM$ scalar estimation problems \cite{Ke_TSP'19}, as
\vspace{-1mm}
\begin{equation}\label{Eq:AMP_Decouple}
{\bf Y} = {\bf S}{\bf X} + {\bf N} \to C_{k,m}^q = x_{k,m} + n_{k,m}^q, \forall k,m,
\vspace{-1mm}
\end{equation}
where $C_{k,m}^q \sim {\cal CN}\left( C_{k,m}^q; x_{k,m}, D_{k,m}^q \right)$ is the equivalent measurement of $x_{k,m}$ obtained in the $q$-th AMP iteration, and $n_{k,m}^q \sim {\cal CN}\left( n_{k,m}^q; 0, D_{k,m}^q \right)$ denotes the effective noise.
Under Bayesian framework, estimating $x_{k,m}$ is to acquire the posterior mean of $x_{k,m}$.
Hence, representing the model (\ref{Eq:RS_Cloud}) by a bipartite graph, the posterior distributions of $x_{k,m}, \forall k,m$ are approximated as
\vspace{-1.5mm}
\begin{equation}\label{Eq:Post_Decouple}
\begin{aligned}
p\left(x_{k,m}|{\bf Y}\right) &\approx p\left( x_{k,m}|C_{k,m}^q, D_{k,m}^q \right)\\
                              &\approx \frac{1} {\widetilde Z} p_0\left( x_{k,m} \right) {\cal CN}\left( x_{k,m}; C_{k,m}^q, D_{k,m}^q\right),
\end{aligned}
\end{equation}
where $p_0\left(x_{k,m}\right)$ denotes the a priori distribution of $x_{k,m}$, and ${\widetilde Z}$ is the normalization factor.
In (\ref{Eq:Post_Decouple}), $D_{k,m}^q$ and $C_{k,m}^q$ are updated at the variable nodes of the bipartite graph as
\vspace{-1.5mm}
\begin{align}
D_{k,m}^q &= \left[\sum\nolimits_g \frac{\left|s_{g,k}\right|^2}{\sigma + V_{g,m}^q}\right]^{-1}, \label{Eq:Var_Update1} \\
C_{k,m}^q &= {\hat x_{k,m}^q} + D_{k,m}^q\sum\nolimits_g{\frac{s_{g,k}^*\left(y_{g,m} - Z_{g,m}^q\right)}{\sigma + V_{g,m}^q}}, \label{Eq:Var_Update2}
\end{align}
where $V_{g,m}^q $ and $Z_{g,m}^q$ are updated at the factor nodes of the bipartite graph as
\vspace{-1mm}
\begin{align}
V_{g,m}^q &= \sum\nolimits_k{\left|s_{g,k}\right|^2{v_{k,m}^q}}, \label{Eq:Fac_Update1}\\
Z_{g,m}^q &= \sum\nolimits_k{s_{g,k}{\hat x_{k,m}^q} - \frac{V_{g,m}^q}{\sigma + V_{g,m}^{q - 1}}\left(y_{g,m} - Z_{g,m}^{q - 1}\right)}. \label{Eq:Fac_Update2}
\end{align}
Moreover, in this paper, we consider a  flexible spike and slab a priori distribution, i.e.,
\begin{equation}\label{Eq:A_Pri}
\begin{aligned}
\!\!\!p_0\left( {\bf X} \right) &= \prod\limits_{m=1}^{M} \prod\limits_{k=1}^{K} p_0\left( x_{k,m} \right)\\
                                &= \prod\limits_{m=1}^{M} \prod\limits_{k=1}^{K} \left[ (1-\gamma_{k,m})\delta(x_{k,m}) + \gamma_{k,m}f(x_{k,m}) \right],\!\!\!
\end{aligned}
\end{equation}
which can well match the actual distribution of channel matrix $\bf X$.
In (\ref{Eq:A_Pri}), $0 < \gamma_{k,m} <1$ is the sparsity ratio, i.e., the probability of $x_{k,m}$ being non-zero, $\delta\left(\cdot\right)$ is the Dirac delta function, and the widely used Gaussian a priori distribution for the channel gains is adopted, i.e., $f\left(x_{k,m}\right) = {\cal CN}\left(x_{k,m}; \mu, \tau\right)$ \cite{Ke_TSP'19}.
By exploiting this a priori model in (\ref{Eq:Post_Decouple}), the posterior distribution of $x_{k,m}$ is obtained as
\begin{equation}\label{Eq:Post_Approx}
\begin{aligned}
\!\!\! p\left(x_{k,m}|C_{k,m}^q, D_{k,m}^q\right) &= \left(1-\pi_{k,m}^q\right)\delta\left(x_{k,m}\right) \\
                                                  &+ \pi_{k,m}^q{\cal CN}\left(x_{k,m}; A_{k,m}^q, B_{k,m}^q\right), \!\!\!
\end{aligned}
\end{equation}
where
\begin{align}
A_{k,m}^q &= \frac{\tau{C_{k,m}^q} + \mu{D_{k,m}^q}}{D_{k,m}^q + \tau},\; B_{k,m}^q = \frac{\tau{D_{k,m}^q}}{\tau + D_{k,m}^q}, \label{Eq:Var_Mean_Var} \\
\pi_{k,m}^q &= \frac{\gamma_{k,m}}{\gamma_{k,m} + \left(1 - \gamma_{k,m}\right)\exp\left( - {\cal L}\right)}, \label{Eq:Belief_Indicator} \\
{\cal L} &= \frac{1}{2}\ln{\frac{D_{k,m}^q}{D_{k,m}^q + \tau}} + \frac{\left|C_{k,m}^q\right|^2}{2D_{k,m}^q} - \frac{\left|C_{k,m}^q - \mu\right|^2}{2\left(D_{k,m}^q + \tau\right)}, \label{Eq:L}
\end{align}
and $\pi _{k,m}^q$ is referred to as the belief indicator.
The posterior mean and variance of $x_{k,m}$ can now be explicitly calculated as
\vspace{-1.5mm}
\begin{align}
g_a\left(C_{k,m}^q, D_{k,m}^q\right) &= {\pi_{k,m}^q}{A_{k,m}^t}, \label{Eq:Post_Mean} \\
g_c\left(C_{k,m}^q, D_{k,m}^q\right) &= {\pi_{k,m}^q}\left(\left|A_{k,m}^q\right|^2 + B_{k,m}^q\right) - \left|g_a\right|^2, \label{Eq:Post_Var}
\end{align}
respectively.

Equations (\ref{Eq:Var_Update1})-(\ref{Eq:Post_Var}) make up the basic steps of MMV-AMP algorithm, while assuming the full knowledge of the sparsity ratio $\gamma_{k,m}, \forall k,m$ and the noise variance $\sigma$, which may be difficult to obtain in practice.
Hence, the expectation maximization algorithm is exploited to learn the unknown hyper-parameters as \cite{Ke_TSP'19}, $\forall k,m$:
\vspace{-1.5mm}
\begin{align}
\gamma_{k,m}^{q+1} &= \pi_{k,m}^{q+1} = \frac{\gamma_{k,m}^q}{\gamma_{k,m}^q + (1-\gamma_{k,m}^q)\exp\left(-{\cal L}\right)}, \label{Eq:EM_Sparse}\\
\sigma_{k,m}^{q+1} &= \frac{1}{G} \! \sum\nolimits_g \!\! {\left[\frac{\left|y_{g,m} \!-\! Z_{g,m}^q\right|^2}{\left|1 \!+\! V_{g,m}^q/\sigma_{k,m}^q\right|^2} + \frac{{\sigma_{k,m}^q}{V_{g,m}^q}}{\sigma_{k,m}^q \!+\! V_{g,m}^q}\right]}. \label{Eq:EM_Noi}
\end{align}

\begin{algorithm}[t]
\caption{MMV-AMP Algorithm}
\label{Alg:MMV-AMP}
\begin{algorithmic}[1]
\REQUIRE Noisy observation ${\bf Y}$, pilot matrix ${\bf S}$, the maximum number of iterations $T_{\rm max}$ and termination threshold $\varepsilon$.
\ENSURE Estimated channel matrix ${\widehat {\bf X}}$ and the related belief indicators $\pi_{k,m}, \forall k,m$.
\STATE $\forall k,m$: Set iteration index $q$ to 1, initialize the hyper-parameters, $\gamma_{k,m}$ and $\sigma_{k,m}$, as in \cite{Ke_TSP'19}, and initialize other parameters as $V_{g,m}^0 = 1$, $Z_{g,m}^0 = y_{g,m}$, $\hat x_{k,m}^1 = 0$, $v_{k,m}^1 = \tau$.
\label{Code:Initial}
\REPEAT
\label{Code:Repeat}
\STATE $\forall g,m$: Update $V_{g,m}^q$ and $Z_{g,m}^q$ according to (\ref{Eq:Fac_Update1}) and (\ref{Eq:Fac_Update2}) at the factor nodes.
\label{Code:Fac_Update}
\STATE $\forall k,m$: Update $D_{k,m}^q$ and $C_{k,m}^q$ according to (\ref{Eq:Var_Update1}) and (\ref{Eq:Var_Update2}) at the variable nodes.
\label{Code:Var_Update}
\STATE $\forall k,m$: Compute the posterior mean and variance as
       \\ ${\hat x_{k,m}^{q+1}} = g_a(C_{k,m}^q, D_{k,m}^q)$, $v_{k,m}^{q+1} = g_c(C_{k,m}^q, D_{k,m}^q)$.
\label{Code:Posterior}
\STATE $\forall k,m$: Update the hyper-parameters $\gamma_{k,m}^{q+1}$ and $\sigma_{k,m}^{q+1}$ as in (\ref{Eq:EM_Sparse}) and (\ref{Eq:EM_Noi}).
\label{Code:Para_Update}
\STATE $\forall k,m$: Refine the update rule for the sparsity ratio,
       \\ $\gamma_{k,m}^{q+1} = \frac{1}{\left|{\cal N}_{k,m}\right|_c} \sum\nolimits_{(o,u) \in {\cal N}_{k,m}} \pi_{o,u}^{q+1}$.
\label{Code:Sparse_Ref}
\STATE $q = q + 1$.
\UNTIL $q > T_{\rm max}$ or $\left\|{\widehat {\bf X}}^q - {\widehat {\bf X}}^{q-1}\right\|_{\rm F}^2 < \varepsilon \left\|{\widehat {\bf X}}^{q-1}\right\|_{\rm F}^2$.
\RETURN ${\widehat {\bf X}}^q$ and $\pi_{k,m}, \forall k,m$.
\end{algorithmic}
\end{algorithm}

In \emph{Algorithm 1}, the sparsity ratio $\gamma_{k,m}$ is the probability that the $\left(k,m\right)$-th element of ${\bf X}$ is non-zero.
In \emph{line \ref{Code:Para_Update}}, $\gamma_{k,m}$ are updated independently for all $k$ and $m$ according to (\ref{Eq:EM_Sparse}), which indicates that the common sparsity pattern described in (\ref{Eq:Sparse_Str}) is not exploited.
To fully exploit the structured sparsity of the channel matrix, we assume that the channel elements associated with the same user have a common sparsity ratio, and further propose to refine $\gamma_{k,m}$ as in \emph{line \ref{Code:Sparse_Ref}}, where we use ${\cal N}_{k,m}\! =\! \left\{\left(o, u\right)|o = k; u = 1, \cdots, M\right\}$.

With the estimate of ${\bf X}$, the estimated active user set and the corresponding channel vectors can be simultaneously acquired.
Specifically, for AUD, we design a user activity detector based on the belief indicators $\pi_{k,m}$, $\forall k,m$, as follows.
We first define a threshold function $r\left(x; \varepsilon_{\rm bi}\right)$, where $r\left(x; \varepsilon_{\rm bi}\right) = 1$ if $ \left|x\right| > {\varepsilon_{\rm th}}$, otherwise $r\left(x; \varepsilon_{\rm bi}\right) = 0$.
\begin{define}\label{Def:BI-AD}
Since the belief indicator $\pi_{k,m}$ tends to be 1 for ${\hat x}_{k,m} \ne 0$ and 0 for ${\hat x}_{k,m} = 0$ after the convergence of MMV-AMP algorithm\footnote{The related proof please refer to \cite{Ke_TSP'19}.}, a belief indicator-based activity detector (BI-AD) is proposed as
\begin{equation}\label{Eq:BI-AD}
{\widehat \alpha_k} = {\widehat \alpha_{b,c}} =\left\{\begin{array}{*{20}{c}}
1, &\frac{1}{M_c}\sum_{m=i+1}^{i+M_c}{r\left(\pi_{k,m}; \varepsilon_{\rm bi}\right) \ge p_{\rm bi}},\\
0, &\frac{1}{M_c}\sum_{m=i+1}^{i+M_c}{r\left(\pi_{k,m}; \varepsilon_{\rm bi}\right) < p_{\rm bi}},
\end{array} \right.
\end{equation}
\end{define}
where $k = (b-1)K_c + c$ and $i = (b-1)M_c + 1$.
We set $\varepsilon_{\rm bi} = 0.5$ to make the missed detection and false alarm probabilities identical, and set $p_{\rm bi} = 0.9$ based on empirical experience\footnote{Given the estimated channel between the $(b,c)$-th user to the $b$-th RRH, if more than $90\%$ of its elements are decided to be non-zero, the $(b,c)$-th UE is declared active.}.
Finally, if the $k$-th user is declared active, its channel to all RRHs is estimated as ${\widehat {\bf X}}_{k,:}$, otherwise, ${\widehat {\bf X}}_{k,:} = {\bf 0}_{1 \times M}$.

\subsection{Fog Computing Based Massive Access}

\begin{figure}[t]
	\centering
	\includegraphics[width=1\columnwidth,keepaspectratio]
    {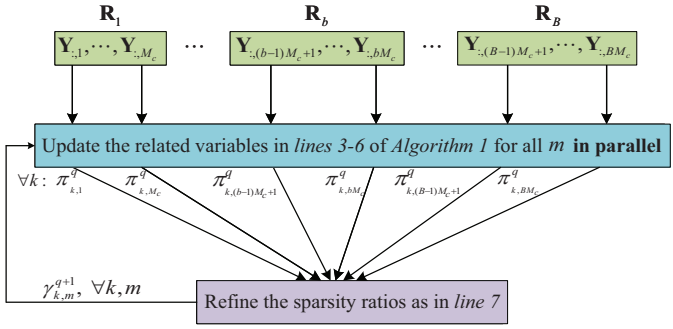}
	\caption{In cloud computing based massive access, the procedure of MMV-AMP algorithm applied to the model (\ref{Eq:RS_Cloud}).}
    \label{Fig:MMV-AMP}
    \vspace{-5mm}
\end{figure}

In this section, based on F-RAN illustrated in Fig. \ref{Fig:F-RAN}, we develop a fog computing deployment for massive access, in which the cloud capabilities are extended closer to the edge of the network.
Specifically, for the whole network, the estimation of user activity and related channels is split, and the corresponding computations are executed at F-APs and fog processing units in close proximity to users.
The split strategy is that each F-AP performs AUD and CE locally based on its own received signal, while several neighboring F-APs cooperate at a fog processing unit for enhanced performance.
The details are summarized in \emph{Algorithm \ref{Alg:Fog_Compute}}.

The difference between C-RAN based and F-RAN based massive access processing is detailed as follows.
For cloud computing paradigm, the detailed procedure of MMV-AMP algorithm applied to the model (\ref{Eq:RS_Cloud}) is summarized in \emph{Algorithm \ref{Alg:MMV-AMP}} and illustrated in Fig. \ref{Fig:MMV-AMP}.
It is clear that, in \emph{lines \ref{Code:Fac_Update}-\ref{Code:Para_Update}} of \emph{Algorithm \ref{Alg:MMV-AMP}}, the signals received at $B$ RRHs are processed in parallel, and only jointly processed in \emph{line \ref{Code:Sparse_Ref}}.
Intuitively, \emph{Line \ref{Code:Sparse_Ref}} leverages the structured sparsity described in (\ref{Eq:Sparse_Str}) to refine the update rule of sparsity ratio $\gamma_{k,m}$, where the channel elements of the same user is assumed to have a common sparsity ratio.
Hence, in fog computing, we propose to replace the RRHs in C-RAN with the F-APs having computation capabilities.
On this basis, for a specific iteration of MMV-AMP algorithm, each F-AP executes the \emph{lines \ref{Code:Fac_Update}-\ref{Code:Para_Update}} locally based on its own received signal, and $\gamma_{k,m}$ in \emph{line \ref{Code:Sparse_Ref}} is jointly refined via F-APs cooperation.

\begin{algorithm}[h]
\caption{Fog Computing Deployment}
\label{Alg:Fog_Compute}
\begin{algorithmic}[1]
\STATE Set iteration index $q$ to 1, and initialize related parameters.
\REPEAT
\STATE $\forall b$: Replacing ${\bf Y}$ with ${\bf R}_b$, the $b$-th F-AP executes \emph{lines 3-6} of \emph{Algorithm 1} locally based on its own received signal ${\bf R}_b$ and known pilot matrix ${\bf S}$.
\label{Code:Update_Parallel}
\STATE $\forall b$: At the $b$-th F-AP, refine the sparsity ratios locally,
       \\ ${\widetilde \pi}_{k,b}^{q+1} = \frac{1}{M_c} \sum\nolimits_{m=(b-1)M_c+1}^{bM_c} \pi_{k,m}^{q+1}$, $\forall k \in [K]$.
\label{Code:Refine_Local}
\STATE $\forall k$: At the fog processing units, jointly refine the sparsity ratios,
       ${\gamma}_{k,m}^{q+1} = \frac{1}{|{\cal B}_k|_c} \sum\nolimits_{b \in {\cal B}_k} {\widetilde \pi}_{k,b}^{q+1}$.
\label{Code:Refine_Joint}
\STATE $q = q + 1$.
\UNTIL $q > T_{\rm max}$.
\STATE $\forall b$: With ${\widehat {\bf X}}_{{\cal K}_b, {\cal M}_b}^q$ and $\pi_{k,m}^q, \forall k \in {\cal K}_b,\forall m \in {\cal M}_b$, the $b$-th F-AP performs AUD and CE for the users in its coverage as in (\ref{Eq:BI-AD}). Here, ${\cal K}_b = \{(b-1)K_c+1, \cdots, bK_c\}$ and ${\cal M}_b = \{(b-1)M_c+1, \cdots, bM_c\}$.
\end{algorithmic}
\end{algorithm}

\begin{figure}[t]
	\centering
	\includegraphics[width=0.75\columnwidth,keepaspectratio]
    {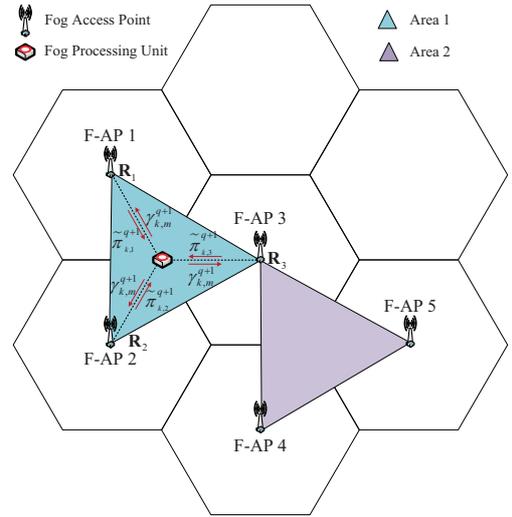}
	\caption{An illustration of F-APs cooperation for fog computing deployment, in which $N_{co} = 3$ is considered.}
    \label{Fig:F-AP_Coop}
    \vspace{-5mm}
\end{figure}

On the other hand, this paper considers a huge size of network to serve a vast area, the channel strengths from a specific active user to far away F-APs are approximate zero because of the large scale fading.
This reveals that, forwarding average belief indicators ${\widetilde \pi}_{k,b}$ obtained at all F-APs to a function node (such as cloud BBU) for jointly refining sparsity ratios, which turns out to be the same effect of \emph{line \ref{Code:Sparse_Ref}} of \emph{Algorithm \ref{Alg:MMV-AMP}}, is not an efficient way.
To make the cooperation of F-APs more flexible, every $N_{co}$ neighboring F-APs are connected to a fog processing unit.
For a specific user surrounded by these F-APs, its sparsity ratios are jointly refined at the associated fog processing unit.
To illustrate the aforementioned F-APs cooperation, as an example, $N_{co} = 3$ is considered in Fig. \ref{Fig:F-AP_Coop}.
The sparsity ratios of users in \emph{Area 1} are jointly refined by \emph{F-APs 1-3}, while that of users in \emph{Area 2} are jointly refined by \emph{F-APs 3-5}.
We describe the F-AP and user association from a user-centric perspective, in which user selects $N_{co}$ closest F-APs for cooperation.
For user $k$, the set of selected F-APs is denoted as ${\cal B}_k \subseteq [B]$ with $|{\cal B}_k|_c = N_{co}$.

\vspace{-0.5mm}
\section{Simulation Results}
For the presented simulation results, we consider a network comprising $B = 7$ hexagonal small cells placed in two tiers.
Each cell contains $K_c = 500$ users uniformly distributed in the coverage with radius $1$ km, among which five percent are active.
Hence, there are in total $K = 3500$ users, and $K_a = 175$ users are active.
Under the C-RAN architecture, we assume one RRH equipped with $M_c$ antennas is located at the center of each cell.
While for F-RAN, the RRHs are replaced with F-APs, and every $N_{co}$ neighboring F-APs are connected to a fog processing unit for cooperation.
The transmit power of users is $23$ dBm, and the background noise power is $-174$ dBm/Hz over $10$ MHz.
For channel model, the complex gain is generated by ${\beta}_{b,k}^l \sim {\cal CN}({\beta}_{b,k}^l; 0,1)$, and the number of MPCs $L$ varies from $8$ to $14$ \cite{Ke_TSP'19}.
Furthermore, $T_{\rm amp} = 200$, $\varepsilon = 10^{-5}$, and the simulation results are obtained by averaging over $3 \times 10^3$ simulation runs.

For performance evaluation, we consider the detection error probability $P_e$ for AUD and the normalized mean square error (NMSE) in dB for CE, which are respectively defined as
\vspace{-1.5mm}
\begin{equation}
P_e = \frac{\sum_k{\left|{\widehat \alpha_k} - {\alpha _k}\right|}}{K},\   {\rm NMSE} =  10{\rm log}_{10}\frac{\left\|{\widehat {\bf X}} - {\bf X}\right\|_{\rm F}^2}{\left\|{\bf X}\right\|_{\rm F}^2}.
\vspace{-1.5mm}
\end{equation}
To verify the superiority of the proposed schemes, a massive access solution based on traditional network architecture is employed as the baseline scheme, where each BS only seeks to detect the users from its own cell while treating the inter-cell interference as noise \cite{Chen_TWC'19}.
Because unless BSs cooperate, it is unlikely that each BS can obtain knowledge about the out-of-cell users.

\begin{figure}[t]
	\centering
	\includegraphics[width=0.7\columnwidth,keepaspectratio]
    {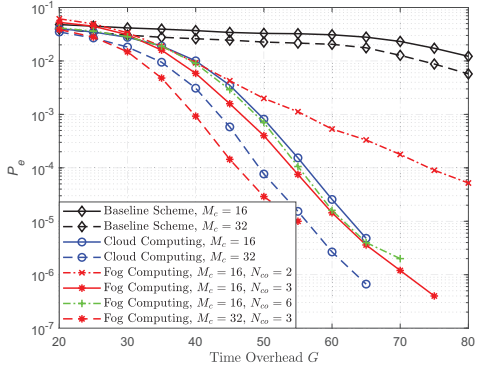}
	\caption{Comparison of detection error probabilities for the proposed massive access schemes and the baseline scheme.}
    \label{Fig:Fig_6}
    \vspace{-5mm}
\end{figure}

\begin{figure}[t]
	\centering
	\includegraphics[width=0.7\columnwidth,keepaspectratio]
    {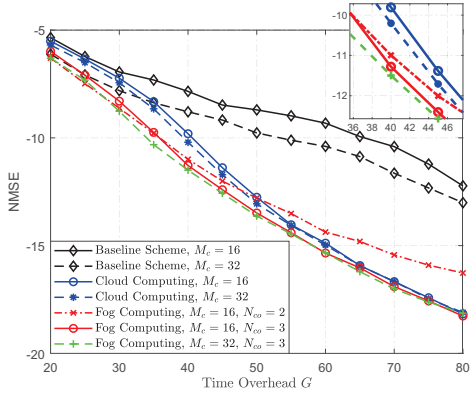}
	\caption{NMSE performance comparison of the proposed schemes and baseline scheme.}
    \label{Fig:Fig_7}
    \vspace{-5mm}
\end{figure}

Fig. \ref{Fig:Fig_6} compares the AUD performance of massive access based on cloud computing and fog computing paradigms, as well as the baseline scheme.
As can be observed, the two proposed schemes outperform the baseline scheme.
Since the inter-cell interference can be recovered via the cooperation of access points.
Moreover, all the three schemes can achieve a better detection performance by equipping more antennas at the access point (RRH, F-AP or BS), as a larger $M_c$ can enhance the structured sparsity observed from multiple receive antennas.

In Fig. \ref{Fig:Fig_6}, for the performance comparison of two processing paradigms, it is clear that, by increasing the number of F-APs for cooperation, i.e., $N_{co}$, the fog computing will approach the performance of cloud computing.
It is worth noticing that fog computing even outperforms cloud computing when $N_{co} = 3$.
This is because the channel strengths from a specific active user to far away access points are approximate zero, it may cause some performance loss by considering these access points when the sparsity ratio is refined as \emph{line \ref{Code:Sparse_Ref}} of \emph{Algorithm \ref{Alg:MMV-AMP}}.
Furthermore, there is a tradeoff between the performance and the cost of practical access point deployment.
The fog computing can reap a faster response time and a better performance while increases the cost of access point deployment.
Fig. \ref{Fig:Fig_7} depicts the CE NMSE performance of the considered schemes, which further demonstrates the superiority of the proposed fog computing based scheme.

\vspace{-2mm}
\section{Conclusion}
This paper investigates the support of grant-free massive access, and proposes two AUD and CE schemes based on cloud computing and fog computing paradigms, respectively.
Considering the cooperation of multiple access points, the proposed schemes can achieve a better performance than the solution based on traditional network architecture.
Furthermore, compared to cloud computing, the fog computing will reap a fast response time and even a better AUD and CE performance, but increase the cost of access point deployment.

\ifCLASSOPTIONcaptionsoff
\newpage
\fi

\end{document}